\documentclass[a4paper,fleqn]{cas-dc}

\usepackage[numbers,sort&compress]{natbib}

\usepackage{graphicx}
\usepackage{amsmath,amssymb}
\usepackage{bm}
\usepackage[dvipsnames]{xcolor}
\usepackage[utf8]{inputenc}
\usepackage{placeins}

\newcommand{\Pup}{\ensuremath{P_{\uparrow}}}
\newcommand{\Pdown}{\ensuremath{P_{\downarrow}}}
\newcommand{\TRutod}{\ensuremath{\Gamma_{\uparrow\to\downarrow}}}
\newcommand{\TRdtou}{\ensuremath{\Gamma_{\downarrow\to\uparrow}}}

\begin{document}
\let\WriteBookmarks\relax
\def\floatpagepagefraction{1}
\def\textpagefraction{.001}
\shorttitle{Quantum spin dynamics of heavy quarks and polarization observables in relativistic heavy-ion collisions}
\shortauthors{ }

\title [mode = title]{Quantum spin dynamics of heavy quarks and polarization observables in relativistic heavy-ion collisions}

\author[1]{Sunil Jaiswal}[orcid=0000-0002-8409-0191]
\ead{jaiswal.61@osu.edu}

\author[2]{Sourav Dey}[orcid=0009-0006-4272-755X]
\ead{sourav.dey@niser.ac.in}

\author[2]{Amaresh Jaiswal}[orcid=0000-0001-5692-9167]
\ead{a.jaiswal@niser.ac.in}


\address[1]{Department of Physics, The Ohio State University, Columbus, Ohio 43210, USA}

\address[2]{School of Physical Sciences, National Institute of Science Education and Research, An OCC of Homi Bhabha National Institute, Jatni-752050, India}


\begin{abstract}
We develop a quantum spin-density-matrix framework for heavy-quark spin dynamics in relativistic heavy-ion collisions. Starting from an initial polarization induced along the magnetic-field direction, we derive the evolution equation for spin polarization within this framework and obtain analytic solutions. The evolved polarization is connected to open heavy-flavor observables via a fragmentation-based hadronization prescription. For vector mesons, the spin-alignment parameter $\rho_{00}$ is constructed by coupling the heavy-quark spin to that of the light antiquark produced during fragmentation. We confront our results with recent ALICE measurements of prompt $D^{*+}$ spin alignment in Pb--Pb collisions at $\sqrt{s_{\rm NN}}=5.02~{\rm TeV}$ and extract an effective depolarization strength that determines the spin-relaxation time scale. Using this fitted parameter, we provide benchmark estimates for $\Lambda_c^+$ and $\bar{\Lambda}_c^-$ polarization, up to an overall spin-transfer normalization. We further estimate the recently proposed elliptic polarization harmonic arising from path-length-dependent depolarization in an anisotropic fireball.
\end{abstract}

\begin{keywords}
Heavy quark  
\sep Magnetic field
\sep Spin density matrix
\sep Spin polarization
\end{keywords}

\maketitle

\section{Introduction}

One of the central goals of the relativistic heavy-ion program is to extract the properties of the QCD medium from experimental observables~\cite{Shuryak:1977ut, Heinz:2013th, JETSCAPE:2020mzn}. This constitutes a challenging inverse problem, as final-state measurements encode the cumulative effects of multiple stages of evolution, including the initial collision geometry, hydrodynamic expansion, hadronization, and hadronic rescattering. The initial geometry sets the conditions for the subsequent fireball evolution, and uncertainties at this early stage can systematically bias the extraction of medium properties from data. A quantitative characterization of the QCD medium therefore requires dedicated probes that retain sensitivity to the initial conditions throughout the entire evolution.

Heavy quarks are natural probes of the QCD medium, as they are predominantly produced in initial hard partonic scatterings and subsequently propagate through the evolving fireball~\cite{Svetitsky:1987gq, Moore:2004tg, vanHees:2005wb, Gubser:2006bz, vanHees:2007me, Akamatsu:2008ge, Das:2010tj, Banerjee:2011ra, Ding:2012iy, das2015toward, STAR:2017kkh, dong2019heavy, ALICE:2021rxa}. Owing to their large masses compared to the typical thermal scale, charm and bottom quarks are not significantly produced during the later stages of the evolution. As a result, they retain memory of the early-time dynamics and serve as valuable probes of the initial fireball geometry, as well as the transport properties, energy loss mechanisms, and hadronization processes of the medium~\cite{Svetitsky:1987gq, Braaten:1991we, GolamMustafa:1997id, Romatschke:2004au, vanHees:2004gq, Bhaduri:2018iwr, Bhaduri:2020lur, Andronic:2021erx, Kumar:2023acr, Andronic:2024oxz}.

In non-central heavy-ion collisions, the initial state also contains strong electromagnetic fields generated by the fast-moving spectator protons ~\cite{Kharzeev:2007jp, Skokov:2009qp, Voronyuk:2011jd, Bloczynski:2012en, Tuchin:2013ie, McLerran:2013hla, Huang:2017tsq, Huang:2022qdn, Jiang:2022uoe}. The magnetic field is largest at very early times and is oriented approximately perpendicular to the reaction plane~\cite{Tuchin:2013ie, McLerran:2013hla, Huang:2022qdn}. Although it decays rapidly, it can still influence particles produced sufficiently early, leading to polarized production. Heavy quarks are particularly well suited in this context, as they are produced in initial hard scatterings while the magnetic field remains appreciable, resulting in polarized initial state. Consequently, the magnetic field can induce a net heavy-quark polarization along its direction~\cite{Dey:2025ail}. The subsequent evolution of this polarization, and its transfer to open heavy-flavor hadron observables, provides a sensitive probe of early-time spin dynamics, including the initial magnetic field and the geometry of the evolving fireball~\cite{Jaiswal:2026ixt}.

Recent measurements of hadron polarization and vector-meson spin alignment have stimulated considerable theoretical activity on spin phenomena in relativistic nuclear collisions ~\cite{ALICE:2025cdf}. In the open-heavy-flavor sector, recent work has explored the possibility that an early magnetic field generates an initial heavy-quark polarization, whose subsequent depolarization can be described through rotational Brownian motion in the QCD medium~\cite{Dey:2025ail}. This provides a useful phenomenological picture for connecting early-time spin polarization with final-state open-heavy-flavor observables. Furthermore, the path-length dependence of depolarization provides a sensitive probe of the initial-state eccentricity, offering an independent window into the early-time collision geometry~\cite{Jaiswal:2026ixt}.

In this Letter, we develop a complementary formulation based on the quantum nature of heavy-quark spin. We describe the heavy-quark polarization by a spin density matrix initialized by the early magnetic field and subsequently depolarized by medium interactions. We focus on the polarization projected along the magnetic-field direction, which is the spin component entering the hadronic observables considered here. The evolved polarization is connected to open-heavy-flavor observables through a fragmentation-based hadronization prescription, which determines how the quark-level spin polarization appears in the final hadronic spin observable. We use this framework to describe $D^{*+}$ spin alignment in Pb--Pb collisions and extract an effective spin-relaxation scale from recent ALICE measurements~\cite{ALICE:2025cdf} at the Large Hadron Collider (LHC). With the fitted depolarization strength, we provide benchmark estimates for $\Lambda_c^+$ and $\bar{\Lambda}_c^-$ polarization. Using the same fitted parameters, we also estimate the recently proposed polarization harmonics, which arise from path-length-dependent depolarization in an anisotropic fireball~\cite{Jaiswal:2026ixt}. We use natural units, $\hbar=c=1$, throughout, and denote three vectors in bold fonts.

\section{Quantum spin dynamics}
\label{sec:spin_dynamics}

In this section, we formulate the dynamics of spin-$1/2$ quarks in terms of a quantum spin density matrix. In general, the spin density matrix may depend on the quark phase-space coordinates and can be written as \cite{Sakurai:1994, Leader_2001}
\begin{equation}
\rho(\tau,\mathbf{x},\mathbf{p}) = \frac{1}{2} \left[ \mathbf{1} + \bm{\mathcal P}(\tau,\mathbf{x},\mathbf{p}) \cdot \bm{\sigma} \right],
\end{equation}
where $\bm{\sigma}$ are the Pauli matrices and $\bm{\mathcal P}(\tau,\mathbf{x},\mathbf{p})$ is the spin-polarization vector. The components of this vector are
\begin{equation}
\mathcal P^i(\tau,\mathbf{x},\mathbf{p}) = \operatorname{Tr} \left[ \rho(\tau,\mathbf{x},\mathbf{p})\sigma_i \right], \qquad i=x,y,z,
\end{equation}
and positivity of the density matrix requires
\begin{equation}
|\bm{\mathcal P}(\tau,\mathbf{x},\mathbf{p})|\leq 1 \,.
\end{equation}
For a given quark trajectory, one may define
\begin{equation}
\rho_\gamma(\tau) \equiv \rho\!\left(\tau,\mathbf{x}(\tau),\mathbf{p}(\tau)\right),
\end{equation}
where $\tau$ denotes the proper time along the trajectory. The spin dynamics can then be described as the time evolution of the density matrix along that path.

In this work, we consider the magnetic field as the source of the initial heavy-quark spin polarization. We therefore choose the quantization axis along the direction of the initial magnetic field and focus on the longitudinal spin polarization along this axis. Restricting the density matrix to the diagonal sector in this basis, the density matrix along a given trajectory reduces to
\begin{equation}
\rho_\gamma(\tau) = \frac{1}{2} \left[ \mathbf{1} + \mathcal P(\tau)\sigma_z \right] =
    \begin{pmatrix}
    \Pup(\tau) & 0 \\
    0 & \Pdown(\tau)
    \end{pmatrix},
\end{equation}
where $\Pup(\tau)$ and $\Pdown(\tau)$ denote the probabilities for the quark spin to be aligned and anti-aligned, respectively, with the chosen quantization axis. They are related to the polarization by
\begin{equation}
\Pup(\tau) = \frac{1+\mathcal{P}(\tau)}{2}\,, \qquad
\Pdown(\tau) = \frac{1-\mathcal{P}(\tau)}{2}\,.
\end{equation}
Equivalently,
\begin{equation}
\Pup(\tau) + \Pdown(\tau)=1, \qquad  \mathcal P(\tau) = \Pup(\tau) - \Pdown(\tau)\,,
\end{equation}
ensuring proper normalization and a consistent definition of polarization. 

Medium interactions can induce spin flips between these two states. Denoting the corresponding transition rates by $\TRutod$ and $\TRdtou$%
    \footnote{Along a given trajectory, the transition rates should be understood as local rates evaluated at $(\tau,\mathbf{x}(\tau),\mathbf{p}(\tau))$. This dependence is suppressed in the notation.}%
, the probabilities obey
\begin{align}
    \frac{d\Pup}{d\tau} &= -\TRutod\, \Pup + \TRdtou\, \Pdown \,,
\label{eq:pup_rate} \\
    \frac{d\Pdown}{d\tau} &= -\TRdtou\, \Pdown + \TRutod\, \Pup \,.
\label{eq:pdown_rate}
\end{align}
The normalization condition is preserved by these equations. Taking the difference of Eqs.~\eqref{eq:pup_rate} and \eqref{eq:pdown_rate}, one obtains the evolution equation for the spin polarization,
\begin{equation}
\frac{d\mathcal{P}}{d\tau} = -\left(\TRutod+\TRdtou\right)\mathcal{P} + \left(\TRdtou-\TRutod \right).
\end{equation}
It is useful to define the spin-relaxation time and equilibrium polarization as
\begin{equation}
\tau_s \equiv \frac{1}{\TRutod+\TRdtou} \,, \qquad
\mathcal{P}_{\rm eq} \equiv \frac{\TRdtou-\TRutod}{\TRutod+\TRdtou}\,.
\label{eq:taus_peq}
\end{equation}
The polarization equation then becomes
\begin{equation}
\frac{d\mathcal{P}}{d\tau} = -\frac{\mathcal{P}-\mathcal{P}_{\rm eq}}{\tau_s}\,.
\label{eq:pol_relaxation}
\end{equation}
For time-dependent rates, the solution is (see Appendix~\ref{app:spin_relaxation})
\begin{equation}
\mathcal{P}(\tau) = \mathcal{P}_0\, D(\tau,\tau_0) + \int_{\tau_0}^{\tau} d\tau'\, \frac{D(\tau,\tau')}{\tau_s(\tau')}\, \mathcal{P}_{\rm eq}(\tau')\,,
\label{eq:pol_general_sol}
\end{equation}
where
\begin{equation}
D(\tau,\tau') \equiv \exp\left[ -\int_{\tau'}^{\tau}\frac{d\tau''}{\tau_s(\tau'')} \right]
\label{eq:depolarization_factor}
\end{equation}
is the polarization damping factor. Here, $\mathcal{P}_0 \equiv \mathcal{P}(\tau_0)$ denotes the initial polarization at the starting proper time $\tau_0$. The parameter $\tau_s$ sets the characteristic relaxation timescale of the polarization, while $\mathcal{P}_{\rm eq}$ specifies the local asymptotic value toward which the spin system evolves.

For constant transition rates, Eq.~\eqref{eq:pol_relaxation} gives
\begin{equation}
\mathcal{P}(\tau) = \mathcal{P}_{\rm eq} + \left(\mathcal{P}_0-\mathcal{P}_{\rm eq}\right)e^{-(\tau-\tau_0)/\tau_s}\,.
\end{equation}
Additionally, if the two transition rates are equal, i.e., $\TRutod=\TRdtou \equiv \Gamma_s$, then $\mathcal{P}_{\rm eq}=0$. Conversely, a vanishing equilibrium polarization implies no preferred spin direction, and therefore equal forward and backward transition rates. In this case, $\tau_s = 1/(2\Gamma_s)$ and the polarization follows a simple exponential decay
\begin{equation}
\mathcal{P}(\tau) = \mathcal{P}_0\,e^{-(\tau-\tau_0)/\tau_s}\,.
\label{eq:pol_const_equal_sol}
\end{equation}
The spin-relaxation description above corresponds to the diagonal sector of the spin-density-matrix evolution in the magnetic-field basis. The early magnetic field prepares an initial polarized spin state, encoded in $\mathcal{P}_0$, while subsequent interactions with the medium reduce this polarization through spin-flip transitions. Note that the above result for evolution of heavy quark spin polarization is consistent with the leading-order result obtained in Ref.~\cite{Dey:2025ail} within a stochastic rotational Brownian motion framework.

\section{Polarization of open heavy-flavor hadrons}
\label{sec:hadron_polarization}

To apply the formalism developed in the previous section for heavy quarks to quantify the polarization of open heavy-flavor hadrons, we must specify the initial polarization $\mathcal{P}_0$, the spin-relaxation timescale $\tau_s$, and the equilibrium polarization $\mathcal{P}_{\rm eq}$ of the light quarks constituting the medium. Experimental measurements of global $\Lambda$ polarization as a function of collision energy show that the polarization decreases with increasing collision energy and becomes small at top RHIC energies~\cite{STAR:2017ckg}. Since the $\Lambda$ baryon is composed of light quarks, this suggests that the bulk medium does not maintain a sizable net spin polarization at high energies. We therefore take the equilibrium heavy-quark polarization to vanish, i.e., $\mathcal P_{\rm eq}=0$, for LHC energies. Consequently, the spin evolution described by Eq.~\eqref{eq:pol_const_equal_sol} provides an adequate description of heavy-quark depolarization. We treat $\mathcal{P}_0$ and $\tau_s$ as phenomenological parameters, to be constrained by comparison with experimental data. Note that the solution in Eq.~\eqref{eq:pol_const_equal_sol} describes the polarization at the heavy-quark level. To connect this to experimentally accessible observables, one must specify how the heavy-quark spin is transferred to the final open heavy-flavor hadron. This mapping is different for baryons and vector mesons, and depends on the relevant hadronization mechanism~\cite{Liang:2004xn, Yang:2017sdk}.

For open heavy-flavor baryons, spin polarization is a key observable. Owing to the large mass of the heavy quark, its spin dynamics can be approximately separated from those of the light degrees of freedom in the hadronic bound state~\cite{Manohar:2000dt}. This justifies treating the evolution of the heavy-quark spin independently, followed by its incorporation into the baryon through hadronization (e.g., quark coalescence). We parametrize the transfer of the evolved heavy-quark polarization to the baryon as
\begin{equation}
\mathcal{P}_{B_Q} = \alpha_{B_Q}\,\mathcal{P}_Q\,,
\label{eq:PBQ}
\end{equation}
where $\alpha_{B_Q}$ is a spin-transfer coefficient encoding the hadronization of the heavy quark into the baryon. This relation will be used later to obtain benchmark estimates for open-charm baryon polarization.

For vector mesons, the observable is instead the spin-alignment parameter $\rho_{00}$, a diagonal element of the spin-1 meson density matrix~\cite{Liang:2004xn, Yang:2017sdk}. It must therefore be constructed at the hadron level by specifying how the heavy quark combines with the light antiquark during hadronization. We describe the spin states of the heavy quark and the light antiquark in the chosen quantization basis as
\begin{equation}
\rho_Q = \frac{1}{2} \left(1+\mathcal{P}_Q\sigma_z\right)\,, \qquad 
\rho_q = \frac{1}{2} \left(1+\mathcal{P}_q\sigma_z\right)\,,
\label{eq:rho_Q_qbar}
\end{equation}
where $\mathcal{P}_Q$ is the heavy quark/anitquark polarization after spin evolution and $\mathcal{P}_q$ denotes the polarization of the light antiquark/quark participating in hadronization. Assuming a factorized spin density matrix for the heavy quark and light antiquark, the combined spin density matrix is given by
\begin{equation}
\rho_{Qq} = \rho_Q\otimes \rho_q\,.
\end{equation}
This combined density matrix can then be projected onto the relevant total-spin states to construct the spin observables of the resulting hadron.

The spin-1 vector meson is obtained by projecting the $Qq$ spin state onto the triplet sector~\cite{Liang:2004xn, Yang:2017sdk},
\begin{align}
|1,1\rangle &= |\uparrow\uparrow\rangle \,, \nonumber\\
|1,0\rangle &= \frac{1}{\sqrt{2}} \left( |\uparrow\downarrow\rangle + |\downarrow\uparrow\rangle \right) \,, \nonumber\\
|1,-1\rangle &= |\downarrow\downarrow\rangle \,.
\end{align}
The spin-alignment parameter is defined as the normalized probability for the $m=0$ triplet state,
\begin{equation}
\rho_{00} = \frac{ \langle 1,0|\rho_{Qq}|1,0\rangle }{ \sum_{m=-1}^{1} \langle 1,m|\rho_{Qq}|1,m\rangle }\,.
\end{equation}
Using Eq.~\eqref{eq:rho_Q_qbar}, one obtains
\begin{align}
\langle 1,0|\rho_{Qq}|1,0\rangle &= \frac{1-\mathcal{P}_Q \mathcal{P}_q}{4}\,, \nonumber \\
\sum_{m=-1}^{1} \langle 1,m|\rho_{Qq}|1,m\rangle &= \frac{3+\mathcal{P}_Q \mathcal{P}_q}{4}\,.
\end{align}
Therefore~\cite{Liang:2004xn, Yang:2017sdk},
\begin{equation}
\rho_{00} = \frac{1-\mathcal{P}_Q\mathcal{P}_q} {3+\mathcal{P}_Q\mathcal{P}_q}\,.
\label{eq:rho00_general}
\end{equation}
Within this factorized construction, the deviation of $\rho_{00}$ from $1/3$ is controlled by the product $\mathcal P_Q\mathcal P_q$. In particular, if the light antiquark is unpolarized, as expected for medium quarks at LHC energies, then $\rho_{00}=1/3$, independent of the heavy-quark polarization. Therefore, a coalescence-based hadronization mechanism involving a heavy quark and an unpolarized light antiquark from the medium is insufficient to explain the large $D^{*+}$ spin alignment observed by the ALICE Collaboration in $5.02~\mathrm{TeV}$ Pb--Pb collisions~\cite{ALICE:2025cdf}.

Further, the recent ALICE measurement of prompt $D^{*+}$ spin alignment in Pb--Pb collisions shows values significantly above $1/3$~\cite{ALICE:2025cdf}. Within the present framework, this behavior can be reproduced only if there exists an effective anticorrelation between the heavy-quark and light-antiquark polarizations. For high-$p_T$ open heavy-flavor mesons, such an anticorrelation can arise naturally in a fragmentation-motivated hadronization prescription. In this picture, the light antiquark is produced during fragmentation, and its spin can be correlated with the spin of the polarized parent heavy quark. This relation can be parametrized as~\cite{Liang:2004xn, Yang:2017sdk}
\begin{equation}
\mathcal{P}_q^{\rm frag} = -\beta\,\mathcal{P}_Q\,,
\label{eq:Pqbar_frag}
\end{equation}
where $\beta$ controls the strength of the spin anticorrelation. Substituting Eq.~\eqref{eq:Pqbar_frag} into Eq.~\eqref{eq:rho00_general}, one obtains
\begin{equation}
\rho_{00}^{\rm frag} = \frac{1+\beta \mathcal{P}_Q^2} {3-\beta \mathcal{P}_Q^2}\,.
\label{eq:rho00_frag}
\end{equation}
For $\beta>0$, this form allows $\rho_{00}>1/3$. For the high-$p_T$ phenomenology considered here, we use the constant equal-rate relaxation limit as a minimal description of polarization loss after the early magnetic field has initialized the spin state. In this limit the medium does not sustain a preferred spin direction, so the equilibrium polarization vanishes and the heavy-quark polarization at hadronization is given by Eq.~\eqref{eq:pol_const_equal_sol}. Substituting the solution for evolution of heavy quark polarization from Eq.~\eqref{eq:pol_const_equal_sol} into Eq.~\eqref{eq:rho00_frag}, we obtain
\begin{equation}
\rho_{00}^{\rm frag}(\Delta\tau) = \frac{1+\beta \mathcal{P}_0^2 e^{-2\Delta\tau/\tau_s}}{3-\beta \mathcal{P}_0^2 e^{-2\Delta\tau/\tau_s}}\,.
\label{eq:rho00_frag_time}
\end{equation}
Here $\Delta\tau$ denotes the effective proper time over which the heavy-quark polarization relaxes before hadronization. 

Following Ref.~\cite{Dey:2025ail}, we estimate this time $\Delta\tau$ by converting an effective in-medium path length in the collision frame to the heavy-quark rest frame. For a heavy quark with transverse-momentum direction $\phi$, this gives
\begin{equation}
\Delta\tau(\phi,p_T,y) = \frac{m_Q \langle L(\phi)\rangle}{|\mathbf{p}|}\,,
\label{eq:delta_tau_phi}
\end{equation}
where $|\mathbf{p}|=\sqrt{p_T^2\cosh^2 y+m_Q^2\sinh^2 y}$, $m_Q$ is the heavy-quark mass, and $y$ denotes the momentum rapidity. Here $\langle L(\phi)\rangle$ is the effective path length for a heavy quark propagating with transverse-momentum direction $\phi$. In the midrapidity region, the dominant variation of the in-medium path length is governed by the azimuthal direction through the transverse fireball geometry. We therefore treat any residual dependence of the effective length on the longitudinal geometry as subleading and absorb it into the effective path-length scale $L_0$, while retaining the explicit kinematic rapidity dependence through $|\mathbf{p}|$. After azimuthal averaging, this gives
\begin{equation}
\Delta\tau(p_T,y) = \frac{m_Q L_0}{|\mathbf{p}|}\,,
\label{eq:delta_tau}
\end{equation}
where $L_0$ is an effective in-medium path-length scale. Using Eq.~\eqref{eq:delta_tau} in Eq.~\eqref{eq:rho00_frag_time}, the spin density matrix for fragmentation can be written as
\begin{equation}
\rho_{00}^{\rm frag}(p_T,y) = \frac{1+A\exp\!\left[-2\kappa m_Q/|\mathbf{p}|\right]}{3-A\exp\!\left[-2\kappa m_Q/|\mathbf{p}|\right]}\,,
\label{eq:rho00_frag_pt_y}
\end{equation}
where $A \equiv \beta \mathcal{P}_0^2$ and $\kappa \equiv L_0/\tau_s$. The parameter $A$ controls the overall magnitude of the spin-alignment, while the dimensionless parameter $\kappa$ characterizes the amount of spin relaxation accumulated over the effective path length.

We compare Eq.~\eqref{eq:rho00_frag_pt_y} with the ALICE measurement of $D^{*+}$ spin alignment in Pb--Pb collisions at $\sqrt{s_{\rm NN}}=5.02~{\rm TeV}$, for the $30$--$50\%$ centrality class and rapidity window $0.3<|y|<0.8$~\cite{ALICE:2025cdf}. To match the experimental binning, we average the theory prediction over the corresponding transverse-momentum and rapidity intervals. More generally, this bin average should be weighted by the differential $D^{*+}$ yield. Since the present analysis does not include a simultaneous description of the $D^{*+}$ spectrum, we use the uniform bin average
\begin{equation}
\left\langle\! \rho_{00}^{\rm frag} \!\right\rangle_{\rm bin} \!\!=\! \frac{1}{\Delta p_T\,\Delta y} \!\int_{p_{T,\min}}^{p_{T,\max}}\! dp_T \!\int_{y_{\min}}^{y_{\max}} \! dy\; \rho_{00}^{\rm frag}(p_T,y)\,,
\label{eq:rho00_bin_average}
\end{equation}
where $\Delta p_T=p_{T,\max}-p_{T,\min}$ and $\Delta y=y_{\max}-y_{\min}$. Since Eq.~\eqref{eq:rho00_frag_pt_y} is even in $y$, the average over $0.3<|y|<0.8$ is evaluated using the positive-rapidity interval. For $D^{*+}$ mesons, we take the heavy-quark mass to be the charm-quark mass, $m_Q=m_c=1.27~{\rm GeV}$.

\begin{figure}
    \centering
    \includegraphics[width=\linewidth]{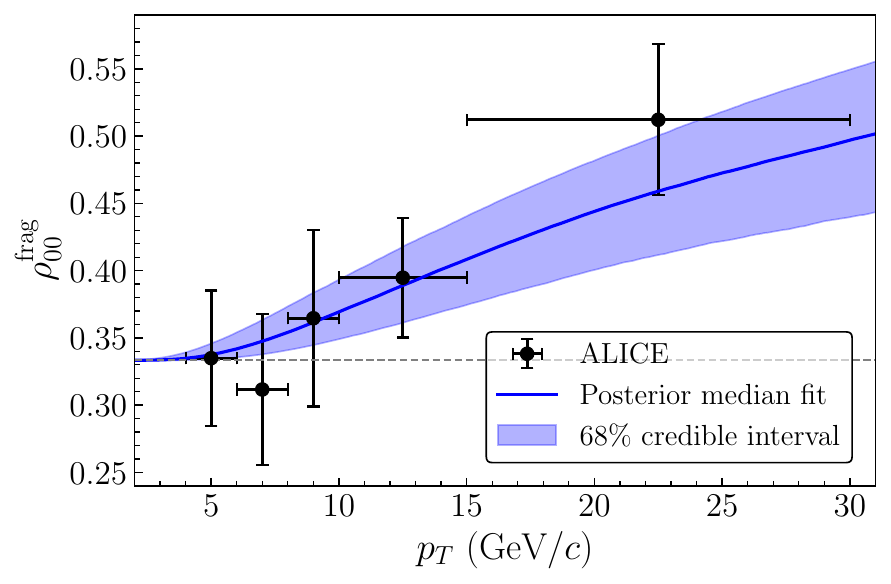}
    \caption{Posterior fit of Eq.~\eqref{eq:rho00_frag_pt_y} to the ALICE measurement of $D^{*+}$ spin alignment~\cite{ALICE:2025cdf}. The model prediction is averaged over the experimental $p_T$ and rapidity bins. The solid curve shows the posterior median and the shaded band shows the $68\%$ credible interval. The dashed horizontal line indicates the unpolarized value $\rho_{00}=1/3$.}
    \label{fig:rho00_fit}
    \vspace{-4mm}
\end{figure}

We then perform a two-parameter Bayesian fit with uniform priors $A\in[0,1]$ and $\kappa\in[0,25]$. The likelihood is constructed using the bin-averaged prediction in Eq.~\eqref{eq:rho00_bin_average} and the experimental uncertainties. Figure~\ref{fig:rho00_fit} shows the posterior predictive result for the spin-alignment observable $\rho_{00}^{\rm frag}$. The solid curve denotes the posterior median, while the shaded band denotes the $68\%$ credible interval obtained by propagating the posterior samples of $A$ and $\kappa$.

The marginalized posterior for the fit parameters favors a sizable spin-alignment amplitude, $A=0.73^{+0.19}_{-0.26}$, together with an effective depolarization strength $\kappa=10.1^{+3.9}_{-3.1}$. Since $\kappa=L_0/\tau_s$, this constraint can be translated into a relaxation time only after specifying an effective path length. For representative choices $L_0=5~{\rm fm}$ and $L_0=10~{\rm fm}$, we obtain $\tau_s=0.50^{+0.22}_{-0.14}~{\rm fm}/c$ and $\tau_s=0.99^{+0.44}_{-0.28}~{\rm fm}/c$, respectively. This extraction absorbs the space-time distribution of heavy-quark production points, path-length fluctuations, possible momentum dependence of the relaxation rate, and the simplified treatment of hadronization into the effective parameters $A$ and $\kappa$. The extracted relaxation scale should therefore be interpreted as a phenomenological estimate of the effective spin-relaxation time relevant for the present kinematic range, rather than as a direct microscopic determination of $\tau_s$.

The same fitted spin-relaxation dynamics can be used to estimate open-charm baryon polarization. For $\Lambda_c^+$, the light degrees of freedom are often modeled as a spin-zero diquark, so that the baryon spin is expected to be strongly correlated with the charm-quark spin. Using Eq.~\eqref{eq:PBQ} and the constant equal-rate relaxation result in Eq.~\eqref{eq:pol_const_equal_sol}, the polarization can be written as
\begin{equation}
\mathcal{P}_{\Lambda_c}(p_T,y) = C_{\Lambda_c} \exp\!\left[-\kappa \frac{m_c}{|\mathbf{p}|}\right],
\label{eq:lambdac_pol}
\end{equation}
where $C_{\Lambda_c}\equiv \alpha_{\Lambda_c}\mathcal P_0$ is an effective normalization that absorbs the unknown initial charm-quark polarization and the spin-transfer coefficient, while the momentum dependence is governed by the fitted depolarization parameter $\kappa$. For the corresponding antibaryon, we get the same magnitude with the opposite sign because $\mathcal P_0\to-\mathcal P_0$, leading to
\begin{equation}
\mathcal{P}_{\bar{\Lambda}_c}(p_T,y) = -C_{\Lambda_c} \exp\!\left[-\kappa \frac{m_c}{|\mathbf{p}|}\right].
\label{eq:antilambdac_pol}
\end{equation}
The opposite signs of the polarization for $\Lambda_c^+$ and $\bar{\Lambda}_c^-$ are consistent with the findings of Ref.~\cite{Dey:2025ail}.

\begin{figure}
    \centering
    \includegraphics[width=\linewidth]{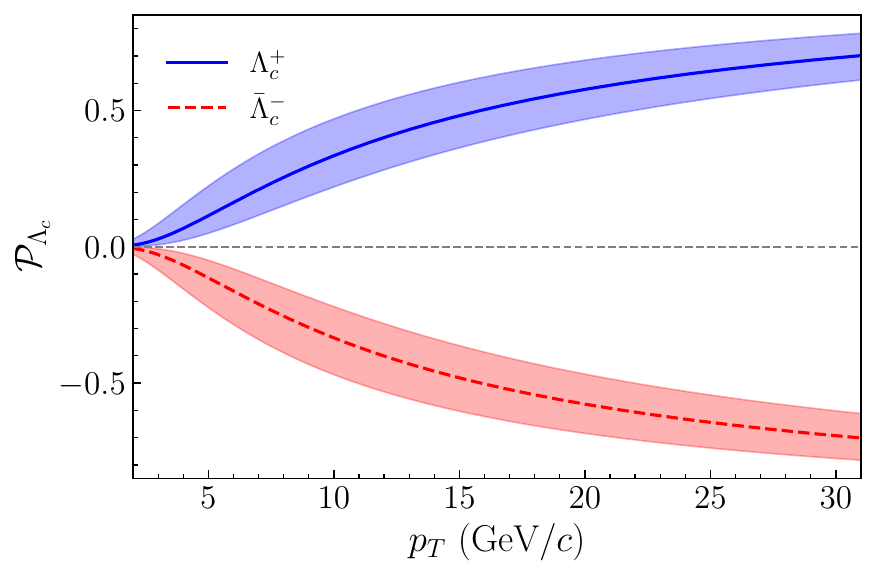}
    \caption{Reference estimate for $\Lambda_c^+$ and $\bar{\Lambda}_c^-$ polarization obtained by using the depolarization parameter $\kappa$ fitted to the ALICE $D^{*+}$ spin-alignment data. The curves use Eqs.~\eqref{eq:lambdac_pol} and \eqref{eq:antilambdac_pol} with $C_{\Lambda_c}=1$. The solid curves denote the posterior medians and the shaded bands denote the $68\%$ credible intervals obtained by propagating the posterior for $\kappa$.}
    \label{fig:lambdac_pol}
    \vspace{-4mm}
\end{figure}

Figure~\ref{fig:lambdac_pol} shows a reference estimate for the polarization of $\Lambda_c^+$ and $\bar{\Lambda}_c^-$ obtained using the depolarization parameter $\kappa$ fitted to the ALICE $D^{*+}$ spin-alignment data ~\cite{ALICE:2025cdf}, with $C_{\Lambda_c}=1$. The baryon-polarization estimates are averaged over the same rapidity window, $0.3<|y|<0.8$. Since $C_{\Lambda_c}$ controls only the overall normalization, results for other choices of the effective spin-transfer amplitude can be obtained by simple rescaling. The magnitude of polarization increases with $p_T$ and approaches its limiting value at high momentum, reflecting the shorter effective proper time available for spin relaxation. The low-$p_T$ behavior should be viewed as a qualitative extrapolation, since additional medium and hadronization effects not included in the present fragmentation-based estimate may become important there.

\section{Polarization harmonics}
\label{sec:polarization_harmonics}

Ref.~\cite{Jaiswal:2026ixt} recently proposed polarization harmonics as a way to characterize the azimuthal structure of spin polarization in an anisotropic fireball. The underlying observation is that the effective path length traversed by a heavy quark depends on its azimuthal direction. Since spin relaxation depends on this path length, the surviving polarization acquires an azimuthal modulation. Here we use the depolarization strength extracted above to estimate the size of these harmonics within the present framework.

In the harmonic estimate, we use the same effective path-length convention as in Eq.~\eqref{eq:delta_tau_phi}. The azimuthal modulation is taken from the path-length result of Ref.~\cite{Jaiswal:2026ixt},
\begin{equation}
\langle L(\phi)\rangle = L_0 \left[ 1 - \sum_{n=2}^{\infty} \frac{\epsilon_n}{n+2} \cos n(\phi-\Psi_n) \right],
\label{eq:path_length_harmonics}
\end{equation}
where $L_0$ is the effective in-medium path-length scale entering Eq.~\eqref{eq:delta_tau}, $\epsilon_n$ denotes the initial eccentricity~\cite{Bhalerao:2020ulk}, and $\Psi_n$ is the corresponding participant-plane angle. Here the azimuthal dependence is controlled by the transverse fireball geometry, while any residual rapidity dependence is already absorbed into the effective scale $L_0$.
Using Eqs.~\eqref{eq:delta_tau_phi} and \eqref{eq:path_length_harmonics}, the azimuth-dependent relaxation exponent becomes
\begin{equation}
\frac{\Delta\tau(\phi,p_T,y)}{\tau_s} = \kappa \frac{m_Q}{|\mathbf p|} \!\left[ 1 - \sum_{n=2}^{\infty} \frac{\epsilon_n}{n+2} \cos n(\phi-\Psi_n) \right],
\label{eq:delta_tau_harmonic}
\end{equation}
where $\kappa=L_0/\tau_s$ is the effective depolarization strength.

We first apply this modulation to the $D^{*+}$ spin-alignment observable. Setting $m_Q=m_c$ and using the azimuth-dependent proper time~\eqref{eq:delta_tau_harmonic} in Eq.~\eqref{eq:rho00_frag_pt_y}, one obtains
{\small 
\begin{equation} 
\rho_{00}^{\rm frag}(\phi,p_T,y) = \frac{ 1\!+\!A\exp\!\left[ -2\kappa \frac{m_c}{|\mathbf p|}\! \left(\! 1 \!-\! \sum_{n=2}^{\infty} \frac{\epsilon_n}{n+2} \cos n\left(\phi-\Psi_n\right)\! \right) \right]}
{3\!-\!A\exp\!\left[ -2\kappa \frac{m_c}{|\mathbf p|}\! \left(\! 1 \!-\! \sum_{n=2}^{\infty} \frac{\epsilon_n}{n+2} \cos n\left(\phi-\Psi_n\right)\! \right) \right] } .
\label{eq:rho00_frag_phi}
\end{equation}
}
For vector mesons, the experimentally relevant spin-alignment signal is the deviation from the unpolarized value $\rho_{00}=1/3$. We therefore introduce
\begin{equation}
\Delta\rho_{00}(\phi,p_T,y) \equiv \rho_{00}^{\rm frag}(\phi,p_T,y)-\frac{1}{3}\,,
\end{equation}
which isolates the spin-alignment signal relative to the unpolarized baseline. Expanding Eq.~\eqref{eq:rho00_frag_phi} to the first order in eccentricities, the azimuthal dependence can be written as
{\small 
\begin{equation}
\Delta\rho_{00}(\phi,p_T,y) = \Delta\rho_{00}^{(0)}(p_T,y) \left[ 1 + 2\sum_{n=2}^{\infty} p_n^{D^{*+}}(p_T,y) \cos n(\phi-\Psi_n) \right],
\label{eq:Delta_rho00_harmonic_def}
\end{equation}
}
where
\begin{equation}
\Delta\rho_{00}^{(0)}(p_T,y) = \frac{4X(p_T,y)}{3[3-X(p_T,y)]}\,, 
\end{equation}
is the isotropic component with
\begin{equation}
X(p_T,y) \equiv A\exp\!\left[-2\kappa \frac{m_c}{|\mathbf p|}\right].
\end{equation}
The corresponding harmonic coefficient is
\begin{equation}
p_n^{D^{*+}}(p_T,y) = \frac{6}{3-X(p_T,y)} \frac{\kappa}{2} \frac{m_c}{|\mathbf p|} \frac{\epsilon_n}{n+2}\,.
\label{eq:Delta_rho00_hg}
\end{equation}
The derivation of the above equations is given in Appendix~\ref{app:D_harmonic}. Equation~\eqref{eq:Delta_rho00_hg} shows that, to leading order in the eccentricities, geometric anisotropies of the fireball generate harmonic modulations of the path-length-dependent spin-alignment signal, with $p_n^{D^{*+}}(p_T,y)\propto\epsilon_n$.

We next apply the same path-length modulation to open-charm baryon polarization. Substituting Eq.~\eqref{eq:delta_tau_harmonic} into Eq.~\eqref{eq:lambdac_pol} gives
{\small
\begin{equation}
\mathcal P_{\Lambda_c}(\phi,p_T,y) = C_{\Lambda_c} \exp\!\left\{\! -\kappa \frac{m_c}{|\mathbf p|} \left[ 1 - \sum_{n=2}^{\infty} \frac{\epsilon_n}{n+2} \cos n(\phi-\Psi_n) \right] \!\right\}\,.
\label{eq:lambdac_pol_phi}
\end{equation}
}
Expanding Eq.~\eqref{eq:lambdac_pol_phi} to first order in the eccentricities, the azimuthal dependence can be written as
{\small
\begin{equation}
\mathcal P_{\Lambda_c}(\phi,p_T,y) = \mathcal P_{\Lambda_c}^{(0)}(p_T,y) \left[ 1 + 2\sum_{n=2}^{\infty} p_n^{\Lambda_c}(p_T,y) \cos n(\phi-\Psi_n) \right],
\end{equation}
}
Here, the isotropic component is
\begin{equation}
\mathcal P_{\Lambda_c}^{(0)}(p_T,y) = C_{\Lambda_c} \exp\!\left[-\kappa \frac{m_c}{|\mathbf p|}\right],
\end{equation}
and the corresponding harmonic coefficient is
\begin{equation}
p_n^{\Lambda_c}(p_T,y) = \frac{\kappa}{2} \frac{m_c}{|\mathbf p|} \frac{\epsilon_n}{n+2}\,.
\end{equation}
The coefficient is independent of the overall amplitude $C_{\Lambda_c}$, which fixes only the polarization normalization. Since the $\bar{\Lambda}_c^-$ polarization differs from $\Lambda_c^+$ only by an overall sign, the same harmonic coefficient is obtained for the antibaryon.

\begin{figure}
    \centering
    \includegraphics[width=\linewidth]{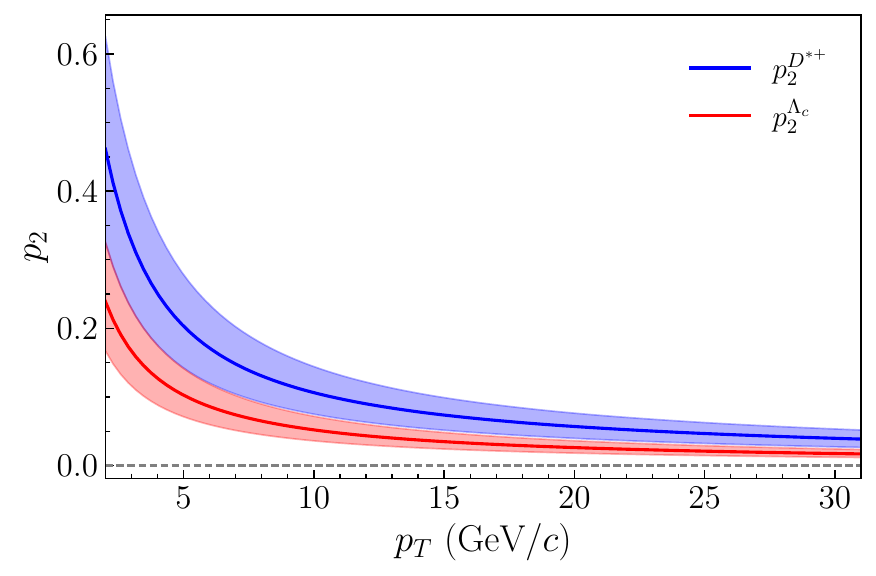}
    \caption{Second polarization harmonic $p_2$ for the $D^{*+}$ spin-alignment observable and for $\Lambda_c^+$ polarization, obtained using the depolarization strength fitted to the ALICE $D^{*+}$ spin-alignment data. The results are averaged over $0.3<|y|<0.8$. The solid curves denote the posterior medians and the shaded bands denote the $68\%$ credible intervals.}
    \label{fig:p2_harmonic}
    \vspace{-4mm}
\end{figure}

Figure~\ref{fig:p2_harmonic} shows the second polarization harmonic $p_2$ as a function of transverse momentum $p_T$ for both $D^{*+}$ and $\Lambda_c^+$. We use $\epsilon_2=0.38$ for the elliptic geometric eccentricity, taken from Monte Carlo Glauber calculations for Pb--Pb collisions at $\sqrt{s_{\rm NN}}=5.02~{\rm TeV}$ in the $30$--$50\%$ centrality class~\cite{Loizides:2017ack}. The estimates are averaged over the rapidity interval $0.3<|y|<0.8$, with uncertainty bands obtained by propagating the posterior samples of $A$ and $\kappa$ for $D^{*+}$ and the same fitted posterior for $\kappa$ for $\Lambda_c^+$. We observe that $p_2$ decreases with increasing $p_T$, indicating that high-$p_T$ charm quarks traverse the medium with minimal depolarization.

\section{Summary and outlook}
\label{sec:summary}

In this Letter, we developed a quantum spin-density-matrix framework for heavy-quark spin relaxation in a QCD medium, building on the picture in which an early-time magnetic field generates an initial heavy-quark polarization~\cite{Dey:2025ail}. The magnetic field is incorporated through the initial spin density matrix, while subsequent interactions with the medium are encoded via spin-flip transition rates. This formulation leads to a relaxation equation for the spin polarization, for which analytic solutions were obtained. The resulting evolution of the heavy-quark polarization is found to be consistent with the leading-order behavior found in Ref.~\cite{Dey:2025ail} within a classical stochastic framework. The heavy-quark polarization was then connected to open-heavy-flavor observables through a fragmentation based hadronization prescription. For vector mesons, the spin-alignment parameter $\rho_{00}$ was constructed at the hadron level by coupling the heavy-quark spin to the light antiquark produced during fragmentation. Comparing the resulting $D^{*+}$ expression with recent ALICE measurements in Pb--Pb collisions at $\sqrt{s_{\rm NN}}=5.02~{\rm TeV}$ ~\cite{ALICE:2025cdf}, we extracted an effective depolarization strength. The same fitted depolarization strength was then used to obtain benchmark estimates for $\Lambda_c^+$ and $\bar{\Lambda}_c^-$ polarization, up to an overall spin-transfer normalization.

We also estimated the recently proposed polarization harmonics generated by path-length-dependent depolarization in an anisotropic fireball. For vector mesons, we considered the harmonic modulation of the spin-alignment signal $\rho_{00}-1/3$, while for open-charm baryons the harmonic was defined directly for the baryon polarization. At leading order in the eccentricities, these harmonics are proportional to the corresponding geometric eccentricities, providing a direct link between spin relaxation and the initial fireball geometry.

The main implication of this framework is that open-heavy-flavor spin observables can serve as clean probes of early-time dynamics. Since heavy quarks are produced before the medium has fully evolved, their polarization can retain sensitivity to the initial magnetic field, while the azimuthal harmonics encode the geometric anisotropy of the produced fireball. Dedicated measurements of open-charm baryon polarization and azimuthal modulations of vector-meson spin alignment would therefore provide complementary access to the early magnetic field, spin relaxation, and the initial geometry of the QCD medium.

A more quantitative description will require replacing the effective spin-flip rates and path-length scale by microscopic calculations of heavy-quark spin relaxation in realistic expanding medium profiles. The initial polarized production of heavy quarks in the early magnetic field should also be treated more microscopically, including its dependence on the space-time structure of the electromagnetic field. A more complete hadronization treatment, including possible recombination effects at low and intermediate $p_T$, will also be important. Together with future measurements, these developments would help establish open-heavy-flavor spin observables as clean probes of early-time electromagnetic fields and the initial fireball geometry.

\section*{Acknowledgments}
S.J. acknowledges support from the CSSI program under Award No.~OAC-2004601 (BAND Collaboration~\cite{BAND_Framework}). S.D. acknowledge financial support from the New Faculty Seed Grant (NFSG), NFSG/PIL/2024/P3825, provided to Arpan Das by the Birla Institute of Technology and Science Pilani, Pilani Campus, India. A.J. gratefully acknowledges Department of Atomic Energy (DAE), India for financial support. 

\appendix 

\section{Time-dependent spin-relaxation}
\label{app:spin_relaxation}

In this appendix, we give the derivation of Eq.~\eqref{eq:pol_general_sol}. We start from the time-dependent relaxation equation~\eqref{eq:pol_relaxation}
\begin{equation}
\frac{d\mathcal{P}}{d\tau} + \frac{1}{\tau_s(\tau)}\,\mathcal{P} = \frac{\mathcal{P}_{\rm eq}(\tau)}{\tau_s(\tau)}\,.
\label{eq:app_pol_time_dep}
\end{equation}
Here $\tau_s(\tau)$ and $\mathcal P_{\rm eq}(\tau)$ are allowed to vary along the heavy-quark trajectory. This time dependence arises because the transition rates can vary as the heavy quark propagates through the evolving medium. Microscopically, the rates may depend on the local medium properties and on the heavy-quark momentum, $\Gamma_{s\to s'}=\Gamma_{s\to s'}(\tau,\mathbf{x}(\tau),\mathbf{p}(\tau))$. Along a specified trajectory, this dependence reduces to an effective proper-time dependence, $\Gamma_{s\to s'}=\Gamma_{s\to s'}(\tau)$.

Equation~\eqref{eq:app_pol_time_dep} is a first-order linear differential equation with time-dependent coefficients and can be solved exactly by the integrating-factor method. We define the damping kernel
\begin{equation}
D(\tau,\tau') \equiv \exp\!\left[ -\int_{\tau'}^{\tau} \frac{d\tau''}{\tau_s(\tau'')} \right].
\end{equation}
It satisfies
\begin{equation}
\frac{\partial}{\partial \tau}D(\tau,\tau') = -\frac{1}{\tau_s(\tau)}D(\tau,\tau'), \quad D(\tau',\tau')=1\,.
\end{equation}
The integrating factor can be written as
\begin{equation}
\mu(\tau) = D^{-1}(\tau,\tau_0) = \exp\!\left[ \int_{\tau_0}^{\tau} \frac{d\tau'}{\tau_s(\tau')} \right].
\end{equation}
Multiplying Eq.~\eqref{eq:app_pol_time_dep} by $\mu(\tau)$, we obtain
\begin{equation}
\frac{d}{d\tau}\left[\mu(\tau)\mathcal{P}(\tau)\right] = \mu(\tau)\,\frac{\mathcal{P}_{\rm eq}(\tau)}{\tau_s(\tau)}\,.
\end{equation}
Integrating from the initial proper time $\tau_0$ to $\tau$ and using $\mathcal{P}(\tau_0)=\mathcal{P}_0$, yields
\begin{equation}
\mu(\tau)\mathcal{P}(\tau)-\mathcal{P}_0 = \int_{\tau_0}^{\tau} d\tau'\, \frac{\mu(\tau')}{\tau_s(\tau')}\, \mathcal{P}_{\rm eq}(\tau')\,.
\end{equation}
Dividing by $\mu(\tau)$ and using $\mu(\tau')/\mu(\tau) = D(\tau,\tau')$ gives the exact solution
\begin{equation}
\mathcal{P}(\tau) = \mathcal{P}_0\, D(\tau,\tau_0) + \int_{\tau_0}^{\tau} d\tau'\, \frac{D(\tau,\tau')}{\tau_s(\tau')}\, \mathcal{P}_{\rm eq}(\tau')\,.
\end{equation}
%

\section{$D^{*+}$ polarization harmonics}
\label{app:D_harmonic}

In this appendix, we present the derivation of Eqs.~\eqref{eq:Delta_rho00_harmonic_def}-\eqref{eq:Delta_rho00_hg} in the main text. Starting from Eq.~\eqref{eq:rho00_frag_phi}, we write the azimuth-dependent
spin-alignment expression as
\begin{equation}
\rho_{00}^{\rm frag}(\phi,p_T,y) = \frac{1+Y(\phi,p_T,y)}{3-Y(\phi,p_T,y)}\,,
\end{equation}
where
\begin{equation}
Y(\phi,p_T,y) = A\exp\!\left[ -2\kappa\frac{m_c}{|\mathbf p|} \left(\! 1 \!-\! \sum_{n=2}^{\infty} \frac{\epsilon_n}{n+2} \cos n(\phi-\Psi_n) \!\right) \right].
\end{equation}
The spin-alignment signal relative to the unpolarized baseline is
\begin{equation}
\Delta\rho_{00}(\phi,p_T,y) \equiv \rho_{00}^{\rm frag}(\phi,p_T,y)-\frac{1}{3} = \frac{4Y(\phi,p_T,y)}{3[3-Y(\phi,p_T,y)]}\,.
\label{eq:app_Delta_exact}
\end{equation}
It is useful to define
\begin{equation}
a(p_T,y) \equiv \kappa\frac{m_c}{|\mathbf p|}, \quad 
S(\phi) \equiv \sum_{n=2}^{\infty} \frac{\epsilon_n}{n+2} \cos n(\phi-\Psi_n)\,,
\label{eq:app_a_S_def}
\end{equation}
so that $Y(\phi,p_T,y) = A e^{-2a} e^{2aS(\phi)}$. For small eccentricities, $S(\phi)=\mathcal{O}(\epsilon_n)$, and therefore
\begin{equation}
Y(\phi,p_T,y) = A e^{-2a} \left[ 1+2a(p_T,y)S(\phi) \right] + \mathcal{O}(\epsilon_n^2)\,.
\end{equation}
Substituting this into Eq.~\eqref{eq:app_Delta_exact} and expanding to first-order in $S(\phi)$, we obtain
\begin{equation}
\Delta\rho_{00}(\phi,p_T,y) = \frac{4Ae^{-2a}}{3(3-Ae^{-2a})} + \frac{8aAe^{-2a}}{(3-Ae^{-2a})^2} S(\phi) + \mathcal{O}(\epsilon_n^2)\,.
\label{eq:app_Delta_expanded}
\end{equation}
Here and below, $a=a(p_T,y)$. The first term is the isotropic component,
\begin{equation}
\Delta\rho_{00}^{(0)}(p_T,y) = \frac{4Ae^{-2a}}{3(3-Ae^{-2a})} = \frac{4X(p_T,y)}{3\left[3-X(p_T,y)\right]}\,,
\end{equation}
where
\begin{equation} 
X(p_T,y) \equiv A e^{-2a} = A\exp\!\left[-2\kappa\frac{m_c}{|\mathbf p|}\right].
\end{equation}
Factoring the isotropic component from Eq.~\eqref{eq:app_Delta_expanded} gives
\begin{equation}
\Delta\rho_{00}(\phi,p_T,y) = \Delta\rho_{00}^{(0)}(p_T,y) \left[ 1 + \frac{6a}{3-X} S(\phi) \right] + \mathcal{O}(\epsilon_n^2)\,.
\end{equation}
Substituting the definition of $a$ and $S(\phi)$ using Eq.~\eqref{eq:app_a_S_def}, we find
{\small 
\begin{equation}
\Delta\rho_{00}(\phi,p_T,y) = \Delta\rho_{00}^{(0)}(p_T,y) \left[ 1 + 2\sum_{n=2}^{\infty} p_n^{D^{*+}}(p_T,y) \cos n(\phi-\Psi_n) \right],
\end{equation}
}
with
\begin{equation}
p_n^{D^{*+}}(p_T,y) = \frac{6}{3-X(p_T,y)} \frac{\kappa}{2} \frac{m_c}{|\mathbf p|} \frac{\epsilon_n}{n+2}\,.
\end{equation}
%

\bibliographystyle{elsarticle-num}
\bibliography{refs}

\end{document}